\documentclass[journal,10pt,twocolumn]{IEEEtran}
\usepackage[colorlinks,bookmarksopen,bookmarksnumbered,citecolor=red,urlcolor=red,]{hyperref}

\usepackage{color}
\usepackage{amsmath,bm} 

\usepackage{amsmath,mathtools}

\usepackage{verbatim}
\usepackage{epsfig,bbm}
\usepackage[colorlinks,bookmarksopen,bookmarksnumbered,citecolor=red,urlcolor=red,]{hyperref}
\usepackage{CJK}
\usepackage{indentfirst}
\usepackage{multirow}
\usepackage{epstopdf}
\usepackage{graphicx}
\usepackage{footmisc}
\graphicspath{{./figures/}}
\usepackage{amsfonts}
\usepackage{mathrsfs}
\usepackage{setspace}
\usepackage{amsmath}
\usepackage{algorithm,algorithmic,amsbsy,amsmath,amssymb,epsfig,bbm,mathrsfs, bbm} %,mathabx
\usepackage{amsthm}
\usepackage{verbatim} 
\usepackage[noadjust]{cite} 
\usepackage[subfigure]{graphfig}

\def \g {\mathbf{g}}

\def \w  {\mathbf{w}}

\def \x {\mathbf{x}}

\def \R {\mathbb{R}}

\usepackage{geometry}
\usepackage{xcolor} 
\allowdisplaybreaks
\usepackage{xcolor} 
\allowdisplaybreaks

\mathtoolsset{showonlyrefs}
\def \g {\mathbf{g}}

\def \w {\mathbf{w}}
\def \x {\mathbf{x}}
\def \xi {\mathbf{x}_i}

\def \Ap {A_\otimes}

\def \R {\mathbb{R}}

\begin{document}

\title{Data Analytics for Fog Computing by Distributed Online Learning with Asynchronous Update\\
\thanks{E-mail addresses: gxli@xidian.edu.cn (G. Li), masonzhao@tencent.com (P. Zhao), lu9@ualberta.ca (X. Lu), jialiu23@stu.xidian.edu.cn (J. Liu), ylshen@mail.xidian.edu.cn (Y. Shen).}
}

%\author{\IEEEauthorblockN{Guangxia Li}
%\IEEEauthorblockA{\textit{School of Computer Science and Technology} \\
%\textit{Xidian University}\\
%Xi'an, China \\
%gxli@xidian.edu.cn}
%\and
%\IEEEauthorblockN{Peilin Zhao}
%\IEEEauthorblockA{\textit{Tencent AI Lab} \\
%\textit{Tencent Inc.}\\
%Shenzhen, China \\
%masonzhao@tencent.com}
%\and
%\IEEEauthorblockN{Xiao Lu}
%\IEEEauthorblockA{\textit{Department of Electrical and Computer Engineering} \\
%\textit{University of Alberta}\\
%Edmonton, Canada \\
%lu9@ualberta.ca}
%\and
%\IEEEauthorblockN{Jia Liu}
%\IEEEauthorblockA{\textit{School of Computer Science and Technology} \\
%\textit{Xidian University}\\
%Xi'an, China \\
%jialiu23@stu.xidian.edu.cn}
%\and
%\IEEEauthorblockN{Yulong Shen}
%\IEEEauthorblockA{\textit{School of Computer Science and Technology} \\
%\textit{Xidian University}\\
%Xi'an, China \\
%ylshen@mail.xidian.edu.cn}
%}

% For over three affiliations, or if they all won't fit within the width
\author{\IEEEauthorblockN{Guangxia Li\IEEEauthorrefmark{1}\IEEEauthorrefmark{2}\IEEEauthorrefmark{3},
Peilin Zhao\IEEEauthorrefmark{4},
Xiao Lu\IEEEauthorrefmark{6}, 
Jia Liu\IEEEauthorrefmark{3} and
Yulong Shen\IEEEauthorrefmark{1}\IEEEauthorrefmark{2}\IEEEauthorrefmark{3}}
\IEEEauthorblockA{\IEEEauthorrefmark{1}Shaanxi Key Laboratory of Network and System Security, Xidian University, China}
\IEEEauthorblockA{\IEEEauthorrefmark{2}State Key Laboratory of Integrated Service Networks, Xidian University, China}
\IEEEauthorblockA{\IEEEauthorrefmark{3}School of Computer Science and Technology, Xidian University, China}
\IEEEauthorblockA{\IEEEauthorrefmark{4}Tencent AI Lab, Tencent Inc., China}
\IEEEauthorblockA{\IEEEauthorrefmark{6}Department of Electrical and Computer Engineering, University of Alberta, Canada}}

\maketitle

\begin{abstract}
Fog computing extends the cloud computing paradigm by allocating substantial portions of computations and services towards the edge of a network, and is, therefore, particularly suitable for large-scale, geo-distributed, and data-intensive applications. 
As the popularity of fog applications increases, there is a demand for the development of smart data analytic tools, which can process massive data streams in an efficient manner.
To satisfy such requirements, we propose a system in which data streams generated from distributed sources are digested almost locally, whereas a relatively small amount of distilled information is converged to a center.
The center extracts knowledge from the collected information, and shares it across all subordinates to boost their performances.
Upon the proposed system, we devise a distributed machine learning algorithm using the online learning approach, which is well known for its high efficiency and innate ability to cope with streaming data.
An asynchronous update strategy with rigorous theoretical support is applied to enhance the system robustness.
Experimental results demonstrate that the proposed method is comparable with a model trained over a centralized platform in terms of the classification accuracy, whereas the efficiency and scalability of the overall system are improved.
\end{abstract}


%-------------------------------------------------------------------
\section{Introduction}
%-------------------------------------------------------------------
\label{sec1}

Recent decades have witnessed an expansion of cloud computing in terms of both service models and applications.
In its simplest form, cloud computing comprises the centralization of computing services using a network of remote servers to enable the sharing of infrastructure resources while achieving economies of scale.
However, this centralization also leads to certain side effects.
Security risks resulting from the exposure of private user data to cloud providers~\cite{lu2018cyber,lu2018managing,lu2018cyber2,niyato2015performance} and service latency incurred by data transmissions represent the two most significant issues.
For privacy- and latency-sensitive applications that require the processing of data in the vicinity of their sources, the cloud's centralized architecture exhibits defects.

Subsequently, an alternative called fog computing has been introduced~\cite{BonomiMZA12,iorga2018fog}. 
In contrast to the cloud, which sends data to a remote location for processing, fog computing allocates substantial amounts of computation, storage, and services toward the edge of a network, i.e., on smart end-devices.
It comprises a large number of fog nodes residing between end-devices and centralized (cloud) services.
Because fog nodes have awareness of their logical locations in the context of the entire system, they are capable of allocating data to apposite locations for processing.
For example, time-sensitive data are analyzed close to their source, and laborious tasks are performed on the cloud.
Fog computing thus can reduce latency, conserve network bandwidth, and to some extent alleviate security problems (because data are less centralized compared with the cloud).
It is suitable for a breed of distributed, latency-aware services and applications, such as Internet-of-things (IoT)~\cite{niyato2015economics}, sensor networks \cite{niyato2016distributed,lu2016sensor}, smart grid systems \cite{LuXiaoPower,niyato2012adaptive,korki2011mac}, cloud systems \cite{li2019data}, communication systems \cite{lu2019intelligent,niyato2015game,lu2019coverage,lu2018ambient}, corporate networks \cite{lu2011payoff,lu2015hierarchical,lu2019ambient,lu2018performance}, and mobile social networks~\cite{zhang2015optimizing}.

With the advent of fog applications, there is a demand for smart data analytic systems with intrinsic distributed real-time processing support.
However, most existing solutions are a stack of off-the-shelf tools, lacking a particular design that caters for the characteristics of fog computing~\cite{YiLL15}.
In this study, we tackle the problem from the core, by generalizing fog data analytics as performing classification over multiple data sources using machine learning.
We present a system in which data streams generated from distributed sources are digested almost locally, whereas a relatively small amount of distilled information is converged to a center.
The center extracts knowledge from the gathered information, and shares it across subordinates.
The subordinates can consist of any computing units, while in a realistic setting, they involve a mass of miniature devices with limited energy, computing power, and communication capacity.

An ideal data analytic algorithm for such a system should have low complexity, high scalability, and a light communication overhead.
We herein adopt an online learning approach for its simplicity and efficiency.
The proposed distributed online multitask learning method employs a master/slave architecture, in which locally calculated gradients and globally updated model vectors are exchanged over the network.
An asynchronous update strategy with rigorous theoretical support is also applied to enhance the system robustness.
Experimental results demonstrate that the classification accuracy of the proposed method is comparable with those of classical models trained in a centralized manner, while the communication overhead is controlled at a reasonable level.
Our approach is suitable for any classification task, and can be ported to any device with moderate computing power to perform data analytics under the fog computing paradigm.

%The rest of this paper is organized as follows.
%Section~\ref{sec2} introduces related work.
%Section~\ref{sec3} provides an overview of the proposed system.
%Section~\ref{sec4} presents our distributed online multitask learning algorithm.
%Section~\ref{sec5} gives experimental results and discussions.
%Section~\ref{sec6} concludes this paper.  

%-------------------------------------------------------------------
\section{Related Work}
%-------------------------------------------------------------------
\label{sec2}

Fog computing has been adopted in a broad range of applications since first being proposed by Cisco in 2012~\cite{BonomiMZA12}.
Augmented reality, online gaming, and real-time video surveillance applications that must process large volumes of data with tight latency constraints are the primary targets for fog computing~\cite{YiLL15}.
Mobile applications running on resource-constrained devices but requiring fast response time, such as wearable assistants~\cite{DubeyYCAYM16} and smart connected vehicles~\cite{HouLCWJC16}, represent an additional use case for fog computing.
Finally, geographically distributed systems represented by wireless sensor networks in general, and smart grids in particular, are compatible with fog computing too.
Additional applications of fog computing, especially those involving big data analytics, are described in~\cite{tang2015hierarchical,ZhuCPNHB13,hong2013mobile}.

Online learning represents a family of efficient algorithms that can construct a prediction model incrementally by processing the training data in a sequential manner, as opposed to batch learning algorithms, which train the predictor by learning the entire dataset at once~\cite{abs-1802-02871}.
On each round, the learner receives an input, makes a prediction using an internal hypothesis that is retained in memory, and subsequently learns the true label.
It then utilizes the new sample to modify its hypothesis according to some predefined rules.
The goal is to minimize the total number of rounds with incorrect predictions.
In general, online learning algorithms are fast, simple, and require few statistical assumptions.
They scale well to a large amount of data, and are particularly suitable for real-world applications in which data arrive continuously.

Existing distributed online learning algorithms can generally be classified into two categories: delayed gradient~\cite{AgarwalD11,recht2011hogwild}, and minibatch gradient~\cite{dekel2012optimal} methods.
The main idea of delayed gradient methods is that workers are allowed to pull the latest model from the master to compute gradients and then send them back, while the master can utilize these gradients to update the model if they are not delayed by too long or sparse enough.
It has been proven that if the number of delayed iterations is not too significant, then the delayed gradient can still converge in line with the standard online gradient method~\cite{AgarwalD11}.
Alternatively, if the gradient is extremely sparse then the delayed gradient method can still converge very effectively~\cite{recht2011hogwild}.
In addition, the main idea behind minibatch gradient methods is to utilize the minibatch technique to reduce the variance of the stochastic gradient estimator, which can in turn improve the convergence rate~\cite{dekel2012optimal}.

%-------------------------------------------------------------------
\section{System Overview}
%-------------------------------------------------------------------
\label{sec3}

\begin{figure}[t]
\centering
\includegraphics[scale=0.43]{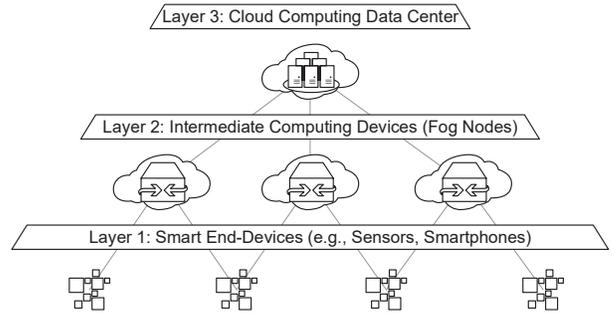}
\caption{Architecture of fog computing.}
\label{fig1}
\end{figure}

We provide an overview of the proposed system from a fog computing perspective.
As shown in Figure~\ref{fig1}, fog computing employs a hierarchical architecture consisting of at least three layers.
Small devices with cost-effective computing powers are located at the edge of the network.
They either act as data sources, by generating data streams of their own, or as data sinks by collecting data from subordinate devices.
Besides gathering data and controlling actuators, they can also perform preparatory data analytics in a timely manner.
The next layer consists of a number of intermediate computing devices, namely fog nodes, each of which is connected to a group of edge devices in the first layer.
These are typically focused on aggregating edge-device data and converting the collected data into knowledge.
The cloud computing data center is in the top layer, providing system-wide monitoring and centralized control.
Such a hierarchy enables the fog to allocate computing resource according to the task scale, thus striking a balance between quick response time and bulk processing power.

In the context of the fog computing architecture, we propose a system that can facilitate data processing in the fog.
It also employs a hierarchical layout, consisting of a generic virtualized device that we call the \textit{Master}, which is dedicated to serving the centralized applications, and numerous client devices that we call \textit{Workers}.
It is assumed that \textit{Workers} are distributed over smart end-devices in different physical locations.
The \textit{Workers} located at the network edge ingest data generated by various sensors, and then transmit the processed information to the \textit{Master}.
Meanwhile, the \textit{Master} sends the global model vector to the \textit{Workers}.
The information flow between the \textit{Master} and \textit{Workers} is bi-directional: \textit{Workers} send locally calculated gradients to the \textit{Master}, and the \textit{Master} sends the global model to the \textit{Workers}.
As with the fog computing paradigm, there are no communications among \textit{Workers}.
It is worth noting that the proposed system resides in layers 1 and 2 of the fog computing architecture.
% Figure 1

There are two important factors to consider when designing such a system.
One is to reduce the data exchange over the network, and the other is to make the system robust when dealing with the inevitable network latency.
As described below, our solution is well-suited to meet these requirements.

%-------------------------------------------------------------------
\section{Proposed Algorithm}
%-------------------------------------------------------------------
\label{sec4}

This section presents a distributed online learning algorithm for binary classification, which serves as the core of the proposed system.
The task for binary classification is to assign new observations into one of the two categories, on the basis of rules that are learned from a training set containing observations whose category memberships are known.
We consider a scenario in which data streams generated from geo-distributed edge devices have to be processed as a coherent whole.
For clarity, it is assumed that a dataset $D$ is distributed over $K$ different devices, and each device is associated with one of $N$ \textit{Workers}, as described above.
As the dataset $D$ is divided into $K$ partitions, we require that the data contained in each partition are homogeneous, e.g., for a sensing application they must be related to a well-defined physical entity that is sensed across different locations.
Therefore, the data from all partitions can be represented in the same global feature space, and it is possible to utilize the shared information between partitions to enhance the overall learning process.
To this end, we can restate our problem as learning from $K$ data sources (or tasks) using $N$ \textit{Workers} under the supervision of one \textit{Master}.

In the following, we employ the notation $I$ to denote the identity matrix.
Given two matrices $M \in \mathbb{R}^{m \times n}$ and $N \in \mathbb{R}^{p \times q}$, we denote the Kronecker product of $M$ and $N$ by $M \otimes N$.
We use $A_\otimes$ as a shorthand for $A \otimes I$.
We will describe the algorithm from the viewpoint of a \textit{Worker} and \textit{Master}, respectively.

%-------------------------------------------------------------------
\subsection{Worker}
%-------------------------------------------------------------------
\label{sec4.1}

In the online learning setting, every \textit{Worker} node observes data in a sequential manner.
Formally speaking, at each step $t$, the $n$-th \textit{Worker} receives a piece of data $(\x_{i_t,t}, y_{i_t,t})$, where $\x_{i_t,t} \in \mathbb{R}^d$ is a $d$-dimensional vector representing the sample, $y_{i_t,t} \in \{-1,1\}$ refers to its class label, and $i_t \in \{1,\ldots,K\}$ denotes the task index (i.e., the index of the task that generated this data).
The classification model for each task is parameterized by a weight vector $\w_{i_t} \in \mathbb{R}^d$.
As there are $K$ tasks involved during learning, we choose to update their weight vectors in a coherent manner.
Specifically, we appoint the \textit{Master} node to maintain a compound vector $\w_t$, which is formed by concatenating $K$ task weights.
That is,
\begin{equation}
\w_t^\top = (\w_{1,t}^\top, \ldots, \w_{K,t}^\top)
\label{eq4.1}
\end{equation}

The model we aim to learn is now parameterized by $\w_t \in \mathbb{R}^{Kd}$.
It is periodically updated on the \textit{Master} side, and distributed to the \textit{Workers} on demand.
Note that we could designate a \textit{Worker} to process a particular task's data the whole time, but this is not compulsory.
Any \textit{Worker} can interact with any task, and vice versa.
We will closely examine how the \textit{Master} updates $\w_t$ later on.

Let us now focus on a single \textit{Worker}.
At time $t$, the \textit{Worker} receives data $(\x_{i_t,t}, y_{i_t,t})$ from the task $i_t$, and the weight vector $\w_t$ from the \textit{Master}.
For ease of presentation, we introduce a compound representation for $\x_{i_t,t}$, and denote the following vector by $\phi_t \in \mathbb{R}^{Kd}$:
\begin{equation}
\phi_t^\top = \left(0, \ldots, 0, \x_{i_t,t}^\top, 0, \ldots, 0 \right)
\label{eq4.2}
\end{equation}

We can formulate the learning process as a regularized risk minimization problem.
To devise the objective function, we first introduce a reproducing kernel Hilbert space (RKHS) $H=\R^{Kd}$ with an inner product $\langle u, v \rangle_H = u^ \top \Ap v$, where $A \in \mathbb{R}^{K \times K}$ is a predefined interaction matrix, which encodes our belief concerning the relationship between the $K$ learning tasks.
Different choices for this interaction matrix result in different geometrical assumptions on the tasks, which will be explained later.

Specifically, for an instance $\x_{i_t,t}$ from the $i_t$-th task, we define the feature map as
\begin{equation}
\Psi(\x_{i_t,t}) = \Ap^{-1} \phi_t
\label{eq4.3}
\end{equation}

Therefore, the kernel product between two instances can be computed as
\begin{equation}
\kappa(\x_{i_s,s},\x_{i_t,t}) = \langle \Psi(\x_{i_s,s}), \Psi(\x_{i_t,t}) \rangle = \phi_s^\top \Ap^{-1}\phi_t
\label{eq4.4}
\end{equation}

If all the training data are provided in advance, then we can formulate the objective as an empirical risk minimization problem in the above RKHS.
That is,
\begin{equation}
\min_{\w} \frac{1}{T} \sum^{T}_{t=1} \log(1 + \exp(-y_{i_t,t} \langle \w, \Psi(\x_{i_t,t}) \rangle_H)) + \frac{\lambda}{2}\| \w \|_H^2
\label{eq4.5}
\end{equation}

However, under the online learning setting, we can only access the $i_t$-th task sample at the $t$-th learning iteration, which can in turn be used to formulate the $t$-th loss:
\begin{equation}
\ell_t(\w_t) = \log(1 + \exp(-y_{i_t,t} \langle \w_t, \Psi(\x_{i_t,t}) \rangle_H)) + \frac{\lambda}{2}\| \w_t \|_H^2
\label{eq4.6}
\end{equation}

For the above loss, we can calculate its gradient with respect to $\w_t$ as follows:
\begin{equation}
\begin{aligned}
\nabla \ell_t(\w_t) 
& = \frac{-y_{i_t,t} \Psi(\x_{i_t,t}) \exp(-y_{i_t,t} \langle \w_t, \Psi(\x_{i_t,t}) \rangle_H}
{1 + \exp(-y_{i_t,t} \langle \w_t, \Psi(\x_{i_t,t}) \rangle_H)} 
+ \lambda \w_t \\
& = \frac{-y_{i_t,t} \Ap^{-1} \phi_t \exp(-y_{i_t,t} \w_t^\top \phi_t)}
{1 + \exp(-y_{i_t,t} \w_t^\top \phi_t)} 
+ \lambda \w_t 
\end{aligned}
\label{eq4.7}
\end{equation}

For the interaction matrix $A$ that encodes our beliefs concerning the relevance between learning tasks, we set it as
\begin{equation}
\begin{aligned}
A = \frac{1}{K} 
\left[\begin{matrix}
a      & -b     & \cdots & -b \\
-b     &  a     & \cdots & -b \\
\vdots & \vdots & \ddots & \vdots \\
-b     & -b     & \cdots & a
\end{matrix} \right]
\end{aligned}
\label{eq4.8}
\end{equation}
where $a = K + b(K-1)$ and $b$ is a user-defined parameter.

It is then easy to verify that
\begin{equation}
\begin{aligned}
A^{-1} = \frac{1}{(1+b)K}
\left[\begin{matrix}
b+K    & b      & \cdots & b \\
b      & b+K    & \cdots & b \\
\vdots & \vdots & \ddots & \vdots \\
b      & b      & \cdots & b+K
\end{matrix} \right]
\end{aligned}
\label{eq4.9}
\end{equation}

Plugging \eqref{eq4.9} into \eqref{eq4.7} and performing some calculations yields
\begin{equation}
\nabla\ell_t(\w_t) = (\g_1, \ldots, \g_j, \ldots, \g_K)
\label{eq4.10}
\end{equation}
with
\begin{equation}
\g_j \!=\!
\begin{cases}
\frac{b+K}{(1+b)K} \frac{-y_{i_t,t} \x_{i_t,t} \exp(-y_{i_t,t} \w_{i_t,t}^\top \x_{i_t,t})}{1 + \exp(-y_{i_t,t} \w_{i_t,t}^\top \x_{i_t,t})} \!+\! \lambda \w_{i_t,t} \hspace{5pt} \textrm{if $j\!=\!i_t$} \\
\frac{b}{(1+b)K}   \frac{-y_{i_t,t} \x_{i_t,t} \exp(-y_{i_t,t} \w_{i_t,t}^\top \x_{i_t,t})}{1 + \exp(-y_{i_t,t} \w_{i_t,t}^\top \x_{i_t,t})} \!+\! \lambda \w_{j,t}   \hspace{6pt} \textrm{otherwise}
\end{cases}
\label{eq4.11}
\end{equation}

It can be observed from \eqref{eq4.10} and \eqref{eq4.11} that the gradient $\nabla\ell_t(\w_t)$ can be computed on a task-wise basis.
The gradient under a multitask setting is composed of the gradients for single tasks with different weights.
Regarding the weights, we can observe the following:
1) The weight for the $i_t$-th task is the largest, while the weights for the other tasks are the same.
2) The parameter $b$ is employed to trade off the differences between the weights.

So far, we have described how a \textit{Worker} derives the gradient using the latest $\phi_t$ (or equivalently, $\x_{i_t,t}$), $y_{i_t,t}$, $\w_t$, and $A$.
Once we have obtained the latest gradient, it seems natural to transmit it to the \textit{Master} immediately to update the model.
However, to reduce network traffic and computational cost incurred by rapid updates, we choose to perform the transmission periodically.
We allocate every \textit{Worker} a buffer of size $m$, to record up to the $m$ latest data samples, and calculate the average gradient whenever the buffer is full.
Specifically, the average gradient of the $n$-th \textit{Worker} is calculated as
\begin{equation}
\frac{1}{m} \sum_{s \in B} \nabla \ell_s(\w_t)
\label{eq4.12}
\end{equation}
where $m$ is the user-defined buffer size and $B$ is the set of indexes for the $m$ buffered examples.
We can control the degree of lazy update by tuning $m$.

In practice, however, we choose not to transmit the result of \eqref{eq4.12} directly over the network.
Referring to \eqref{eq4.7}, we can decompose \eqref{eq4.12} as
\begin{equation}
\frac{1}{m} \sum_{s \in B} \nabla\ell_s(\w_t) = 
 \Ap^{-1}\bar{\g} + \lambda \w_t
\label{eq4.13}
\end{equation}
where
\begin{equation}
\bar{\g} = 
\frac{1}{m}\sum_{s \in B} \frac{-y_{s_t,t} \phi_s \exp(-y_{s_t,t} \w_t^\top \phi_s)}{1 + \exp(-y_{s_t,t} \w_t^\top \phi_s)}
\label{eq4.14}
\end{equation}

The $\bar{\g}$ in \eqref{eq4.14} is what the \textit{Worker} actually computes and transmits to the \textit{Master}.
The \textit{Master} will utilize the received $\bar{\g}$ and the task-relationship matrix $A$ to construct the average gradient $\frac{1}{m} \sum_{s\in B} \nabla \ell_s(\w_t)$.
The reason for this is that $\bar{\g}$ can be more sparse than $\frac{1}{m} \sum_{s\in B} \nabla \ell_s(\w_t)$, especially when $K$ is large.
Transmitting a sparse vector rather than a dense one can reduce the network cost.
Note that the sparsity results from two factors: 1) most blocks of $\phi_s$ are zero, and 2) we choose to shift the Kronecker product operation, which can reduce the sparsity of resulting vector, to the \textit{Master} side.

%-------------------------------------------------------------------
\subsection{Master}
%-------------------------------------------------------------------
\label{sec4.2}

The \textit{Master} node employs the gradient information provided periodically by the \textit{Workers} to update $\w_t$, and then sends the updated $\w_t$ to the \textit{Workers} whenever requested.
Specifically, the $n$-th \textit{Worker} transmits the $\bar{\g}$ in \eqref{eq4.14} to the \textit{Master}.
The \textit{Master} utilizes the received $\bar{\g}$ to compute the average gradient as in \eqref{eq4.13}.

To counter the network latency, we let the \textit{Master} to record the outage durations of each \textit{Worker}, i.e., $\tau_n, n\in\{1,\ldots,N\}$, where $\tau_n$ denotes the number of learning rounds in which the $n$-th \textit{Worker's} $\bar{\g}$ has not been utilized for the model update. 
At the beginning of each learning round, the \textit{Master} will first check whether the largest outage value $\max \tau_n$ exceeds the allowed threshold $\tau$.
If so, then the \textit{Master} will choose that $\bar{\g}$ to update the model (it may have to wait a short time for the corresponding \textit{Worker} to response).
Otherwise, the \textit{Master} will use the latest $\bar{\g}$ from any \textit{Worker} to update the model.
This strategy is known as the delayed gradient descent approach~\cite{AgarwalD11}.
It can help to improve the convergence rate of a distributed online learning algorithm.

Finally, we summarize the pseudocode for \textit{Worker} and \textit{Master} in Algorithm~\ref{alg1} and \ref{alg2}, respectively.
% Algorithm 1
\begin{algorithm}[h]
	\caption{Distributed Online Multitask Learning (Worker)}
	\label{alg1}
	\begin{algorithmic}[1]
		\STATE {\bf Input}: a sequence of instances $(\x_{i_t,t},y_{i_t,t})$, $i_t\in[K]$, $t\in[T]$; a parameter $m$ specifying the buffer size
		\STATE {\bf Output}: a vector $\bar{\g}$ conveying the average gradient information
		\STATE {\bf Initialize}: $\w_0 = \mathbf{0}$
		\STATE {Receive $m$ instances regardless of which task they belong to}
		\STATE {Pull the latest model $\w$ from the master}
		\STATE {Compute the average gradient $\bar{\g}$ for $m$ examples from the online stream according to \eqref{eq4.14}}
		\STATE {Transmit $\bar{\g}$ to the Master}
	\end{algorithmic}
\end{algorithm}
\vspace{-10pt}
% Algorithm 2
\begin{algorithm}[h]
	\caption{Distributed Online Multitask Learning (Master)}
	\label{alg2}
	\begin{algorithmic}[1]
		\STATE {\bf Input}: a regularization parameter $R$; a parameter $\tau$ specifying the maximum outage allowed; a $K \times K$ interaction matrix $A$; a number of gradients $\bar{\g}_n, n=1,\ldots,N$ provided by the Workers
		\STATE {\bf Output}: a vector $\w$ conveying the learned model
		\STATE {\bf Initialize}: set $\tau_n=0$ for $n=1,\ldots,N$, $\w_0 = \mathbf{0}$
		\FOR {$t=1,\ldots,T$}
		\STATE {$i = \mathop{\mathrm{arg}} \max_n \tau_n$}
		\IF {$\tau_i = \tau$}
		\STATE {Wait for $\bar{\g}_i$ from the Worker $i$}
		\STATE {Update $\w_t = \w_{t-1} - \eta_t [ A_\otimes^{-1} \bar{\g}_i + \lambda \w_{t-1-\tau_i}]$}
		\STATE {Set $\tau_i=0$ and $\tau_n=\tau_n+1$, $\forall n \neq i$}
		\ELSE
		\STATE {Receive the latest $\bar{\g}_j$ from any Worker $j$ who responses first}
		\STATE {Update $\w_t = \w_{t-1} - \eta_t [ A_\otimes^{-1} \bar{\g}_j + \lambda \w_{t-1-\tau_j}]$}
		\STATE {Set $\tau_j=0$ and $\tau_n=\tau_n+1$, $\forall n \neq j$}
		\ENDIF
		\STATE {$\w_{t+1} \leftarrow min(1,R/ \left\| \w_t \right\|) \w_t$}
		\ENDFOR
	\end{algorithmic}
\end{algorithm}

%-------------------------------------------------------------------
\section{Experimental Results}
%-------------------------------------------------------------------
\label{sec5}

We evaluate the proposed algorithm using a synthetic dataset, first introduced in~\cite{sheldon2008graphical}.
The problem is to discriminate two classes in a two-dimensional plane with nonlinear decision boundaries.
By $\mathbf{x}=(x_1,x_2)$, we denote a point in two-dimensional space.
The basic classification boundaries are generated according to the rule $g(\mathbf{x};\mathbf{a}) = \mathrm{sign}(x_2 - h(x_1;\mathbf{a}))$, where $h(x;\mathbf{a})$ is a family of nonlinear functions consisting of the first few terms of a Fourier series, defined as $h(x;\mathbf{a}) = a_1 \sin(x - a_0) + a_2 \sin(2(x - a_0)) + a_3 \cos(x - a_0) + a_4 \cos(2(x - a_0))$.
A rotation is applied to the decision boundary to create multiple tasks.
Let $R_\theta$ denote the operator that rotates a vector by $\theta$ radians in a counterclockwise direction about the origin.
The final family of classifiers is $f(\mathbf{x};\mathbf{a},\theta) = g({R_{\theta}}\mathbf{x};\mathbf{a})$ with $\theta$ as an additional parameter.
% Figure 2
\begin{figure}[t]
\centering
\subfigure [A task] {
\label{synthetic-task-1}
\centering
\includegraphics[scale=0.3]{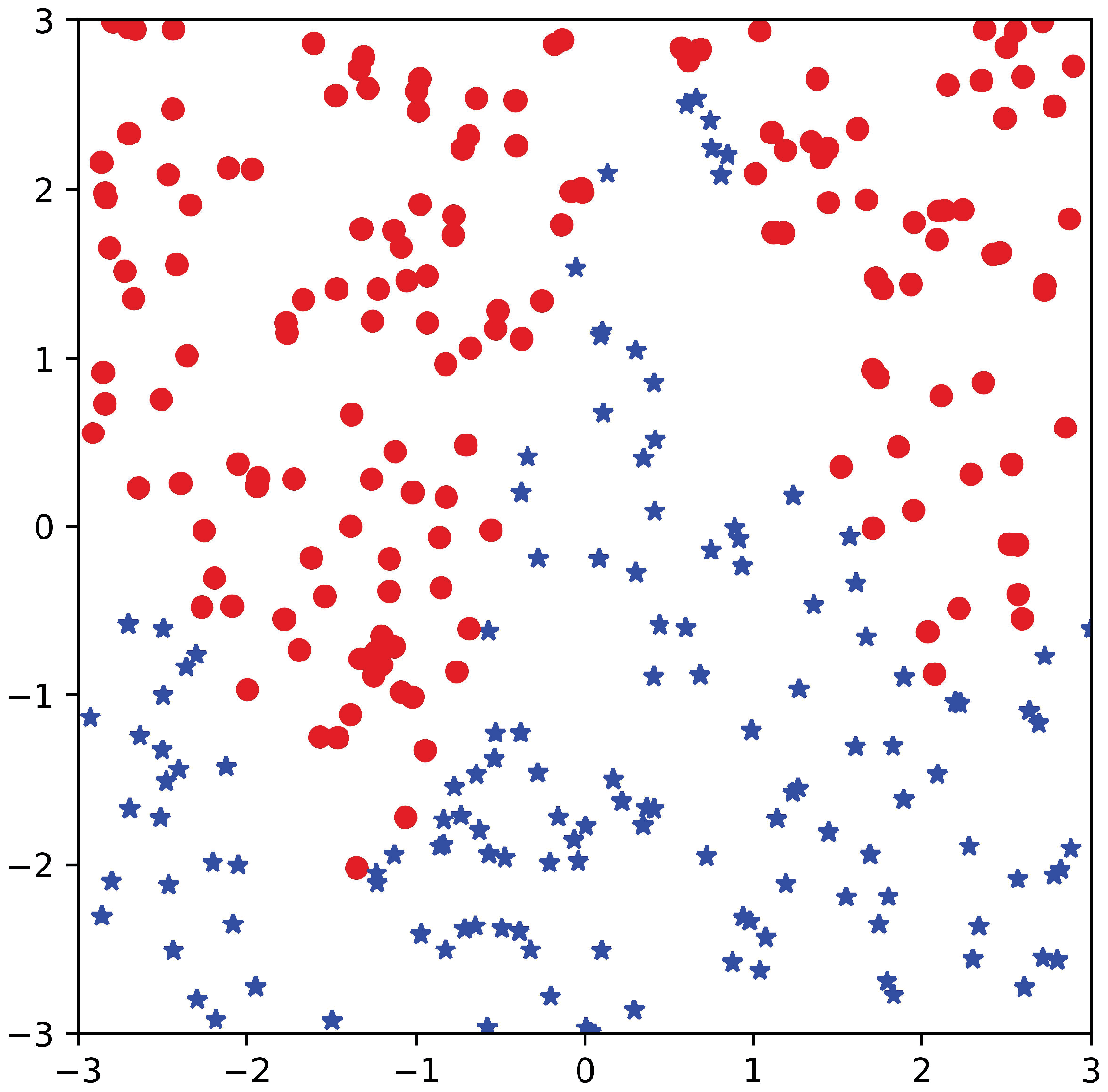}}
\centering
\subfigure [Another task] {
\label{synthetic-task-2}
\centering
\includegraphics[scale=0.3]{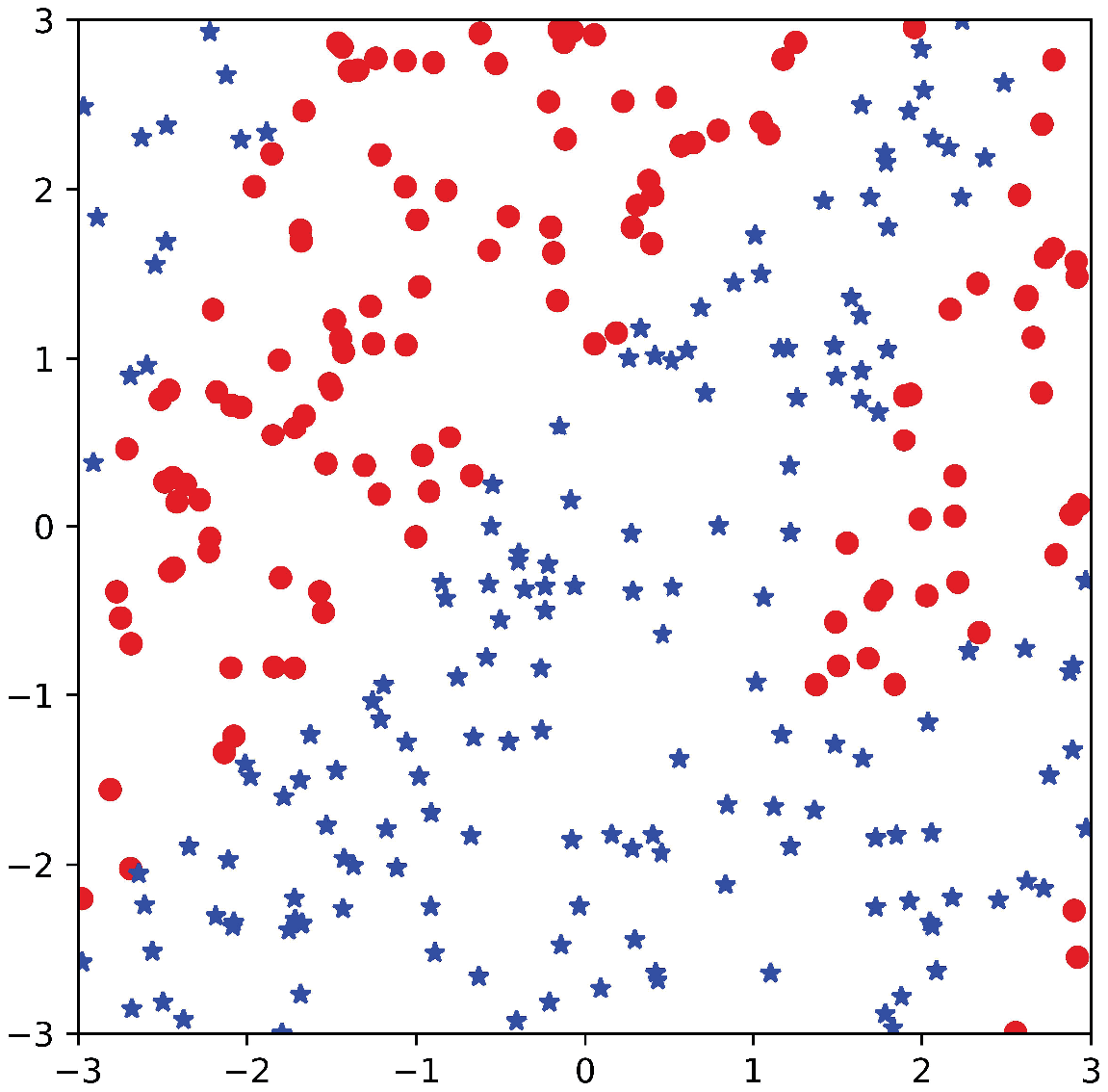}}
\caption{Two tasks selected from the synthetic dataset. Each task contains 300 samples. The two classes are represented by circles and crosses.}
\label{fig2}
\end{figure}

A total of 64 tasks are involved in the experiment.
Their parameters are generated via a random walk in a parameter space with Gaussian increments.
The initial values are set as $\mathbf{a}^{(1)} = (0,1,1,1,1)$ and $\theta^{(1)} = 0$.
For $t=2,\ldots,64$, $\mathbf{a}^{(t)} = \mathbf{a}^{(t-1)} + {\epsilon}_t, {\epsilon}_t \sim N(0,\sigma^2 I)$ and $\theta^{(t)} = \theta^{(t-1)} + \delta_t, \delta_t \sim N(0,\sigma^2(\pi/4)^2)$.
The parameter $\sigma$ controls the step size, and hence the task similarity.
We set it to 0.3.
A training sample is generated by choosing an input $\mathbf{x}$ uniformly at random from the square $x_1, x_2 \in [-3, 3]$, and then labeling it according to $f(\mathbf{x};\mathbf{a},\theta)$ (see Figure~\ref{fig2} for an example).
Because the problem is not linearly separable, we add seven additional features, which are derived from the original $x_1$ and $x_2$ via a mapping $(x_1,x_2) \mapsto (x_1,x_2,x_1x_2,x_1^2,x_2^2,x_1^3,x_2^3,x_1x_2^2,x_1^2x_2)$.

\begin{figure}[t]
\centering
\includegraphics[scale=0.37]{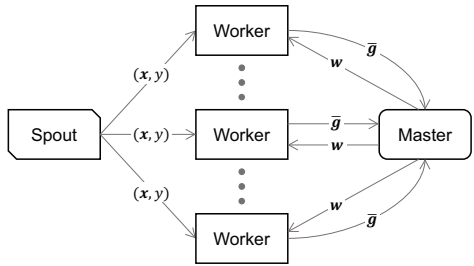}
\caption{Architecture of the simulation system. The \textit{Spout} acts as a streaming data source, by sending task-related data $(\x,y)$ to a randomly selected \textit{Worker}. The \textit{Worker} delivers the computed $\bar{\g}$ to the \textit{Master}, and receives the updated model $\w$ in return.}
\label{fig3}
\end{figure}

We construct a simulation system in accordance with the fog computing paradigm.
Its underlying implementation consists of a set of PC-hosted programs communicating with each other in an asynchronous, full-duplex mode.
We employ \textit{asyncoro}, a Python library for asynchronous, concurrent, and distributed programming, as the framework~\cite{WEB:http://asyncoro.sourceforge.net}.
As illustrated in Figure~\ref{fig3}, the system imitates a streaming data source using a \textit{Spout} node.
It continues to produce the aforementioned synthetic data, and sends them to a randomly selected \textit{Worker}.
However, in reality, the data source could be Twitter status updates, a branch of temperature sensors, or any entities that continuously generate data.
Situations are commonly encountered in real-world applications in which there exist thousands of smart devices (e.g., sensors or smartphones), which are analogous to the \textit{Spout} node in this simulation.

\begin{figure}[h]
\centering
\includegraphics[scale=0.4]{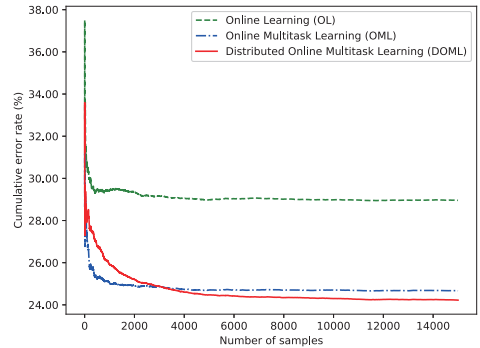}
\caption{Variations of the cumulative error rate, averaged over 64 tasks along the entire online learning process.}
\label{fig4}
\end{figure}

Our experiment involves 64 tasks (data sources), eight \textit{Workers}, and a single \textit{Master}.
We set the learning rate $\eta=0.01$, regularization parameter $\lambda=0.001$, and interaction matrix parameter $b=6$, by referring to a small validation set.
We employ the \textit{cumulative error rate} as an evaluation metric, which is given by the ratio of the number of mistakes made by the online learner to the number of samples received to date.
Besides the proposed distributed online multitask learning (DOML) algorithm, we include two existing methods for comparison.
The first is online multitask learning (OML), which adopts a similar multitask learning approach to DOML, but runs on a single machine.
The other is a vanilla online learning (OL) method, which maintains a single model for all tasks.
Figure~\ref{fig4} depicts the variations in the cumulative error rate averaged over 64 tasks along the entire online learning process.
Note that although online learning algorithms are capable of dealing with infinite samples, we truncate the result by 15,000 samples per task, as the curve will become flat before that, indicating that the model has reached a stable state.
% Figure 3
% Figure 4

It can be observed from Figure~\ref{fig4} that the two multitask learning methods (DOML and OML) achieve the lowest cumulative error rates, demonstrating that they are effective for learning problems with a commonly shared representation across multiple related tasks.
The difference between DOML and OML is marginal (24.22\% vs. 24.67\% in terms of the cumulative error rate evaluated at the 15,000-\textit{th} epoch).
However, owing to the distributed architecture, DOML enjoys more efficiency and almost unlimited horizontal scalability compared to the standalone OML.
In our experimental setting with an Intel Core i7 2.4 GHz CPU and 8 GB RAM, DOML configured with eight \textit{Workers} is able to process hundreds of thousands of samples within a few seconds.
Furthermore, it is obvious that such processing power can easily be increased by introducing more \textit{Workers} into the system.

\begin{figure}[t]
\centering
\includegraphics[scale=0.4]{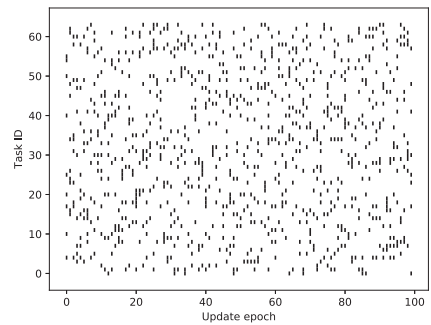}
\caption{Illustration of the sparsity of the gradient vector $\bar{\g}$ delivered from a \textit{Worker} to the \textit{Master} during the first 100 learning epochs, with the buffer size set to 10. A black spot denotes a nonzero element, whereas blank areas are all zeros.}
\label{fig5}
\end{figure}

Next, we analyze the communication cost of DOML.
Regarding the cost related to data sources and \textit{Workers} (i.e., the data emitted from the \textit{Spout} in this experiment), it is clear that any dataset will be divided into $N$ chunks and distributed to $N$ \textit{Workers}.
This makes every \textit{Worker's} load equal to $1/N$ of the original problem load.
This is especially helpful when devices with moderate computing power encounter a massive dataset that exceeds any of their processing capacities.
Regarding the information exchange between \textit{Workers} and the \textit{Master}, a straightforward implementation would involve $N$ \textit{Workers} periodically sending the \textit{Master} their gradient information, calculated by averaging the gradients for $m$ samples.
However, as described in Section~\ref{sec4.1}, we choose to defer the calculation of the average gradient to the \textit{Master} side, so that we can utilize the sparsity of $\bar{\g}$, as in \eqref{eq4.14}, to save bandwidth.
Given $N$ \textit{Workers} learning from $K$ data sources (or tasks), with the buffer size set as $m$, the \textit{Master} has to maintain $N$ communication channels, each of which conveys a sparse vector with only $m/K$ entities with nonzero values.
As illustration, we depict the occurrences of nonzero elements of the delivered gradient vector $\bar{\g}$ corresponding to the first 100 learning epochs in Figure~\ref{fig5}.
% Figure 5

It is noteworthy that the performance of the vanilla OL algorithm is inferior to those of DOML and OML (28.96\% in Figure~\ref{fig4}).
Our intuition is that learning related tasks via a single model is inappropriate, as this ignores the individual task characteristics.
To verify this, we adjust the parameter $\sigma$ to generate a set of more similar tasks and a set of less similar tasks.
For a dataset with $\sigma$ set as 0.1, the 64 tasks are more similar to each other, making a single model adequate for all of them.
This is verified by the experimental results: 18.71\% (OL) vs. 21.34\% (DOML) in terms of the cumulative error rate.
In contrast, by setting $\sigma$ to 0.5, the increased inconsistencies between tasks cause the OL error rate increase to 41.93\%, whereas DOML achieves a lower value of 27.05\%.
Thus, it is obvious that compared with OL, DOML is more suitable for real-world applications where data are not strictly homogeneous.

%-------------------------------------------------------------------
\section{Conclusion}
%-------------------------------------------------------------------
\label{sec6}

In this paper, we have proposed the use of online machine learning to classify streaming data in a manner that is compatible with the fog computing paradigm.
To cope with a large number of edge devices and large volumes of data for real-time low-latency applications, we devised a distributed online multitask learning algorithm, which fits well with the architecture of fog systems.
The experimental results demonstrated that jointly learning multiple related tasks is superior to a single model working in a standalone mode.
More importantly, the classification accuracy of the proposed method is comparable with that of a centralized algorithm trained over the entire dataset, while the efficiency is enhanced and the network overhead is reduced.
For future work, we aim to extend our experiments to a more substantial scale and additional applications.
In conclusion, our work serves as an initial attempt to develop low-latency, real-time and online data analytic tools for fog computing.

\section*{Acknowledgment}

The work was supported by the National Natural Science Foundation of China (Grant No.:~61602356, U1536202), the Key Research and Development Program of Shaanxi Province, China (Grant No.:~2018GY-002), and the Shaanxi Science \& Technology Coordination \& Innovation Project (Grant No.:~2016KTZDGY05-07-01).

 %\bibliographystyle{IEEEtran}
 %\bibliography{mybibliography}

\begin{thebibliography}{99}

\bibitem{lu2018cyber}
X. Lu, D. Niyato, H. Jiang, P. Wang, and H. V. Poor, “Cyber insurance
for heterogeneous wireless networks,” IEEE Communications Magazine,
vol. 56, no. 6, pp. 21–27, 2018.


\bibitem{lu2018managing}
X. Lu, D. Niyato, N. Privault, H. Jiang, and P. Wang, “Managing physical
layer security in wireless cellular networks: A cyber insurance approach,”
IEEE Journal on Selected Areas in Communications, vol. 36, no. 7, pp.
1648–1661, 2018.


\bibitem{lu2018cyber2}
 X. Lu, D. Niyato, N. Privault, H. Jiang, and S. S. Wang, “A cyber insurance
approach to manage physical layer secrecy for massive mimo cellular
networks,” in 2018 IEEE International Conference on Communications
(ICC). IEEE, 2018, pp. 1–6.

\bibitem{niyato2015performance}
D. Niyato, P. Wang, D. I. Kim, Z. Han, and L. Xiao, “Performance
analysis of delay-constrained wireless energy harvesting communication
networks under jamming attacks,” in 2015 IEEE Wireless Communications
and Networking Conference (WCNC). IEEE, 2015, pp. 1823–1828.

\bibitem{BonomiMZA12}
F. Bonomi, R. A. Milito, J. Zhu, and S. Addepalli, “Fog computing and its
role in the internet of things,” in Proceedings of the first edition of the MCC
workshop on Mobile cloud computing, MCC@SIGCOMM 2012, 2012, pp.
13–16.

\bibitem{iorga2018fog}
 M. Iorga, L. Feldman, R. Barton, M. J. Martin, N. S. Goren, and
C. Mahmoudi, “Fog computing conceptual model,” Tech. Rep. 325, 2018.

\bibitem{niyato2015economics}
D. Niyato, X. Lu, P. Wang, D. I. Kim, and Z. Han, “Economics of
internet of things (iot): An information market approach,” arXiv preprint
arXiv:1510.06837, 2015.

\bibitem{niyato2016distributed}
 ——, “Distributed wireless energy scheduling for wireless powered sensor
networks,” in 2016 IEEE International Conference on Communications
(ICC). IEEE, 2016, pp. 1–6.

\bibitem{lu2016sensor}
X. Lu, “Sensor networks with wireless energy harvesting.” 2016.

\bibitem{LuXiaoPower}
 X. Lu, D. Niyato, and P. Wang, “1 power management for wireless base
station in smart grid environment: Modeling and optimization.”

\bibitem{niyato2012adaptive}
 D. Niyato, X. Lu, and P. Wang, “Adaptive power management for wireless
base stations in a smart grid environment,” IEEE Wireless Communications,
vol. 19, no. 6, pp. 44–51, 2012.

\bibitem{korki2011mac}
M. Korki, H. L. Vu, C. H. Foh, X. Lu, and N. Hosseinzadeh, “Mac
performance evaluation in low voltage plc networks,” ENERGY, pp. 135–
140, 2011.

\bibitem{li2019data}
 G. Li, P. Zhao, X. Lu, J. Liu, and Y. Shen, “Data analytics for fog computing
by distributed online learning with asynchronous update,” in ICC 2019-
2019 IEEE International Conference on Communications (ICC). IEEE,
2019, pp. 1–6.

\bibitem{lu2019intelligent}
 X. Lu, E. Hossain, T. Shafique, S. Feng, H. Jiang, and D. Niyato, “Intelligent
reflecting surface (IRS)-enabled covert communications in wireless
networks,” arXiv preprint arXiv:1911.00986, 2019.

\bibitem{niyato2015game}
D. Niyato, P. Wang, D. I. Kim, Z. Han, and L. Xiao, “Game theoretic
modeling of jamming attack in wireless powered communication networks,”
in 2015 IEEE International Conference on Communications (ICC). IEEE,
2015, pp. 6018–6023.

\bibitem{lu2019coverage}
 X. Lu, E. Hossain, H. Jiang, and G. Li, “On coverage probability with
type-ii harq in large-scale uplink cellular networks,” IEEE Wireless Communications
Letters, 2019.

\bibitem{lu2018ambient}
X. Lu, D. Niyato, H. Jiang, D. I. Kim, Y. Xiao, and Z. Han, “Ambient
backscatter assisted wireless powered communications,” IEEE Wireless
Communications, vol. 25, no. 2, pp. 170–177, 2018.

\bibitem{lu2011payoff}
X. Lu, P. Wang, and D. Niyato, “Payoff allocation of service coalition in
wireless mesh network: A cooperative game perspective,” in 2011 IEEE
Global Telecommunications Conference-GLOBECOM 2011. IEEE, 2011,
pp. 1–5.

\bibitem{lu2015hierarchical}
 ——, “Hierarchical cooperation for operator-controlled device-to-device
communications: A layered coalitional game approach,” in 2015 IEEE
Wireless Communications and Networking Conference (WCNC). IEEE,
2015, pp. 2056–2061.

\bibitem{lu2019ambient}
 X. Lu, D. Niyato, H. Jiang, E. Hossain, and P. Wang, “Ambient backscatter assisted
wireless-powered relaying,” IEEE Transactions on Green Communications
and Networking, 2019.

\bibitem{lu2018performance}
 X. Lu, G. Li, H. Jiang, D. Niyato, and P. Wang, “Performance analysis
of wireless-powered relaying with ambient backscattering,” in 2018 IEEE
International Conference on Communications (ICC). IEEE, 2018, pp. 1–6.

\bibitem{zhang2015optimizing}
Y. Zhang, D. Niyato, P. Wang, and X. Lu, “Optimizing content relay
policy in publish-subscribe mobile social networks,” in 2015 IEEE Wireless
Communications and Networking Conference (WCNC). IEEE, 2015, pp.
2167–2172.

\bibitem{YiLL15}
 S. Yi, C. Li, and Q. Li, “A survey of fog computing: Concepts, applications
and issues,” in Proceedings of the 2015 Workshop on Mobile Big Data,
Mobidata@MobiHoc 2015, 2015, pp. 37–42.

\bibitem{DubeyYCAYM16}
 H. Dubey, J. Yang, N. Constant, A. M. Amiri, Q. Yang, and K. Mankodiya,
“Fog data: Enhancing telehealth big data through fog computing,” CoRR,
vol. abs/1605.09437, 2016.

\bibitem{HouLCWJC16}
 X. Hou, Y. Li, M. Chen, D. Wu, D. Jin, and S. Chen, “Vehicular fog
computing: A viewpoint of vehicles as the infrastructures,” IEEE Trans.
Vehicular Technology, vol. 65, no. 6, pp. 3860–3873, 2016.

\bibitem{tang2015hierarchical}
 B. Tang, Z. Chen, G. Hefferman, T. Wei, H. He, and Q. Yang, “A
hierarchical distributed fog computing architecture for big data analysis
in smart cities,” in Proceedings of the ASE BigData \& SocialInformatics
2015. ACM, 2015, p. 28.

\bibitem{ZhuCPNHB13}
 J. Zhu, D. S. Chan, M. S. Prabhu, P. Natarajan, H. Hu, and F. Bonomi,
“Improving web sites performance using edge servers in fog computing architecture,”
in Seventh IEEE International Symposium on Service-Oriented
System Engineering, 2013, pp. 320–323.

\bibitem{hong2013mobile}
K. Hong, D. Lillethun, U. Ramachandran, B. Ottenw¨alder, and B. Koldehofe,
“Mobile fog: A programming model for large-scale applications
on the internet of things,” in Proceedings of the second ACM SIGCOMM
workshop on Mobile cloud computing. ACM, 2013, pp. 15–20.

\bibitem{abs-1802-02871}
S. C. H. Hoi, D. Sahoo, J. Lu, and P. Zhao, “Online learning: A
comprehensive survey,” CoRR, vol. abs/1802.02871, 2018.

\bibitem{AgarwalD11}
 A. Agarwal and J. C. Duchi, “Distributed delayed stochastic optimization,”
in Advances in Neural Information Processing Systems 24, 2011, pp. 873–
881.

\bibitem{recht2011hogwild}
B. Recht, C. Re, S. Wright, and F. Niu, “Hogwild: A lock-free approach
to parallelizing stochastic gradient descent,” in Advances in neural information
processing systems, 2011, pp. 693–701.

\bibitem{dekel2012optimal}
 O. Dekel, R. Gilad-Bachrach, O. Shamir, and L. Xiao, “Optimal distributed
online prediction using mini-batches,” Journal of Machine Learning Research,
vol. 13, no. Jan, pp. 165–202, 2012.

\bibitem{sheldon2008graphical}
D. Sheldon, “Graphical multi-task learning,” in NIPS’08 Workshop on
Structured Input and Structured Output, 2008.

\bibitem{li2019data}
G. Pemmasani, asyncoro: Asynchronous, Concurrent, Distributed
Programming with Python, 2016. [Online]. Available:
http://asyncoro.sourceforge.net
\end{thebibliography}

\end{document}